\documentclass[10pt]{article}
\usepackage[cp1251]{inputenc}
\newcommand{\sss}{\scriptstyle}
\usepackage[dvips]{graphicx}

\def\lsim{\
  \lower-1.2pt\vbox{\hbox{\rlap{$<$}\lower5pt\vbox{\hbox{$\sim$}}}}\ }
\def\gsim{\
  \lower-1.2pt\vbox{\hbox{\rlap{$>$}\lower5pt\vbox{\hbox{$\sim$}}}}\ }

\begin{document}
\title{Some mechanisms of  ``spontaneous'' polarization of superfluid He-$4$}
\author{\textbf{Maksim D. Tomchenko}
\bigskip \\ {\small Bogoliubov Institute for Theoretical Physics} \\
 {\small 14b, Metrolohichna Str., Kyiv 03680, Ukraine}
 \\ {\small E-mail: mtomchenko@bitp.kiev.ua}}
 \date{\empty}
 \maketitle
 \large
 \sloppy
    \textit{Previously, a quantum ``tidal''  mechanism of polarization of the atoms of He-II was proposed,
    according to which, as a result of interatomic interaction,
    each atom of He-II acquires small fluctuating dipole and multipole moments, oriented chaotically on the average.
    In this work, we show that, in the presence of a temperature or density gradient in He-II,
    the originally chaotically oriented tidal dipole moments of the atoms become
    partially ordered, which results in volume polarization of He-II. It is found that the
    gravitational field of the Earth induces electric induction ${\sss \triangle}\varphi \sim 10^{-7}\,V$
    in He-II (for vessel dimensions of the order of
    $10\,cm$). We study also the connection of polarization and acceleration, and discuss
    a possible  nature of the electric signal ${\sss \triangle} \varphi  \approx k_{B}{\sss \triangle}
    T/2e$ observed by A.\,S.\,Rybalko in experiments with second
    sound. }
%


       \section{INTRODUCTION}

In a series of fine experiments, A.S. Rybalko, E.Ya. Rudavskii,
S.P. Rubets, \textit{et al}. obtained a number of interesting
results testifying that the atoms of superfluid $He^4$ possess
electric properties \cite{r1}--\cite{r-sht}. In studies of both a
standing second-sound half-wave in He-II \cite{r1} and
 torsional oscillations of a film of He-II \cite{r2},
 the alternating electric voltage $U$ synchronous, respectively,
 to the second sound and torsional oscillations was observed. This
  voltage was not related to external electromagnetic fields
 (in experiments \cite{r1}, the external voltage was present and is supplied to a heater, but its frequency is twice less than that
of the observed signal $U$)
  and can be a consequence of the volume polarization of He-II. The effects \cite{r1,r2} were not explained up to now, though the attempts to elucidate
the experiment with
second sound \cite{r1} were made in a number of works
\cite{kos}--\cite{gl}.

    Free atoms of $He^4$ create no electric field far from themselves, because they have zero charge and
    zero dipole and multipole moments. However,
    an atom of helium, being surrounded by other atoms, can acquire a dipole moment (DM), which follows from
the tidal mechanism \cite{bw1}--\cite{lt1arc,lt1}, by which a DM
is induced by the interaction with neighboring atoms (we call the
mechanism ``tidal'', since the deformation of electron shells
  of atoms in this case reminds gravitational tides).
      In what follows, we
           will show that a dielectric is polarized due to the gradient of concentration $n$ or temperature $T$.
     The effect arises at the consideration of the interaction of atoms. Also we  will
     study the connection of polarization and acceleration.
          The idea of
     inducing the polarization by the concentration gradient was earlier considered
     in \cite{vol84,mel,naz,nac08}. Below, we will carry on a more exact analysis and determine the volume polarization of He-II
     in a second-sound wave.

         \section{POLARIZATION OF He-II DUE TO  A GRAVITATIONAL  FIELD AND THE GRADIENTS OF  DENSITY AND TEMPERATURE}

       It is obvious that a single atom freely falling in a gravitational
       field  $\textbf{g}$ is not polarized, since the gravity force causes
       the same acceleration $\textbf{g}$ of the nucleus and electrons of the atom.

      Consider a dielectric (He-II), being at rest
      in a gravitational field.
           The gravity force acting on every atom of a dielectric in the equilibrium state,
      should be balanced by the difference of interatomic forces which
     act on the given atom
      from the side of neighboring atoms. This means that, in this case, the concentration gradient
       must exist   in the dielectric.
          We now evaluate the polarization of He-II induced by the gravitational
          field.
         We start from the equations of two-fluid hydrodynamics \cite{pat,xal}
  \begin{equation}
 D\textbf{v}_s/Dt=-\nabla\mu +  \textbf{g},
  \label{hd1} \end{equation}
  \begin{equation}
 \rho D\textbf{v}/Dt=-\nabla p +  \rho\textbf{g},
  \label{hd2} \end{equation}
   \begin{equation}
 dp=\rho d\mu+SdT.
  \label{hd3} \end{equation}
  In the absence of macroscopic motions ($ \textbf{v}_n = \textbf{v}_s =0$), we get
    \begin{equation}
 \nabla\mu =
  \textbf{g}, \ \nabla p =  \rho\textbf{g}.
  \label{hd4} \end{equation}
   Let us consider that $\rho=\rho(p,T)$, and $\nabla T =0$ in the stationary case. Then
    \begin{equation}
  \nabla p =  \frac{\partial p}{\partial \rho}|_T \nabla \rho\approx
   \frac{\partial p}{\partial \rho}|_S \nabla \rho =  u_{1}^{2}\nabla \rho.
  \label{hd5} \end{equation}
   Relations (\ref{hd4}) and (\ref{hd5}) allow us to determine $\nabla\rho$ which compensate
   the gravity force:
    \begin{equation}
 \nabla\rho =
    \rho\textbf{g}/u_{1}^{2}.
  \label{hd6} \end{equation}
    We direct the $Z$ axis upward, so that $\textbf{g}=-g
      \textbf{i}_{z}$. Relation (\ref{hd6}) implies that the mean
    distance  between atoms of He-II,
     $\bar{R}$, depends on the height as
    \begin{equation}
   \frac{\partial\bar{R}}{\partial z} = \frac
   {\bar{R}g}{3u_{1}^{2}}.
  \label{hd8} \end{equation}
     Let us position the coordinate origin at the middle of the layer of He-II. Then relation (\ref{hd6}) yields the density distribution for helium
     \begin{equation}
 \rho = \rho_0 e^{-gz/u_{1}^{2}}.
  \label{hd7} \end{equation}
   At $z=0$, $\rho$ is equal to the mean density $\rho_0$ ($0.1452 \mbox{g}/\mbox{cm}^3$ at $T\lsim 1.3\,K$ and the saturated vapor pressure).
    At the height characteristic of experiments, $z=10\,\mbox{cm}$, we have $\rho = \rho_0
    (1-1.6\cdot 10^{-5})$, i.e. the difference between $\rho$ and $\rho_0$
for the real size of a vessel is insignificant.

    We now evaluate the polarization of He-II. It was shown quantum-mechanically in works \cite{bw1,bw2,lt1arc} that each of two quiescent
    interacting atoms of $He^4$, which are located at a distance $R$ from each other, acquires a tidal DM (TDM)
     which is directed by the minus to another atom and equal to
   \begin{equation}
  \textbf{d}_{tid} = -D_{7} |e| \frac{a_{B}^{8}}{R^7} \textbf{n},   \label{da}     \end{equation}
        where $\textbf{n} =\textbf{R}/R$ is the unit vector directed to another atom, and
     $a_B = \frac{\hbar^2}{me^2}=0.529\,\mbox{\AA}$ is the Bohr
     radius. The quantity $D_{7}\approx 18.4$ according to \cite{bw1,bw2}, whereas simpler calculations in \cite{lt1arc}
     give $D_{7}\approx 25.2\pm 2$. Taking into account both of results, we accept
            \begin{equation}
   D_{7} \simeq 23\pm 5.   \label{d7}     \end{equation}
       It is convenient to represent $\textbf{d}_{tid}$ in the form
         \begin{equation}
  \textbf{d}_{tid} = -d_{0}\frac{\bar{R}_{0}^{7}}{R^7}\textbf{n}, \ \ \
       d_{0}= D_{7}|e|\frac{a_{B}^{8}}{\bar{R}_{0}^7} \simeq 1.88\cdot 10^{-5}|e|\mbox{\AA}.   \label{da2}     \end{equation}

     In liquid helium, each atom is surrounded by many other atoms;
therefore, its electronic cloud is subject to many deformations,
and DM is equal to the sum of DMs induced by all its neighbors. In
this case, the total tidal DM  of the atoms is equal to zero in
view of their irregular location. However, if a density or
temperature gradient is present in He~II, TDMs of the atoms become
partially ordered leading to an overall polarization in He~II (see
below).

     Due to the gradient of $\rho$ along $z$, the mean distance $\bar{R}_{1}$ from the
     given atom of He-II to the adjacent lower one is somewhat less than the distance $\bar{R}_{2}$ to the
     adjacent upper atom. According to (\ref{hd8}) and (\ref{da2}),
     each atom of helium acquires the noncompensated mean (over the time or atoms) DM in this case:
                \begin{equation}
  \textbf{d}_{g}  = \textbf{i}_{z} d_{0}\left (\frac{\bar{R}^7_{0}}{\bar{R}_{1}^7}-\frac{\bar{R}_{0}^{7}}{\bar{R}_{2}^7}
    \right ) \approx  -\frac{7d_{0}\bar{R}\textbf{g}}{3u_{1}^{2}}.
         \label{d0}     \end{equation}
               DM (\ref{d0}) of an atom arises due to a difference of the mean TDMs induced by lower and upper atoms.
        But this estimate does not take the randomness of positions of atoms
        in helium into account. The shape \cite{ss} of the binary distribution function
        $g_{c}(r)$ for He-II implies that the distance between atoms of
     He-II varies mainly in the interval $R\simeq (5/6 \div
     7/6)\bar{R}$, and the mean deviation $\delta R = R-\bar{R}$ satisfies the relation
       $(\delta R)^{2} \simeq (\bar{R}/6)^{2}$.
       By introducing the quantity $\delta R$ in (\ref{d0}), we note that its value is different for
       different pairs of atoms:
    \begin{equation}
    R_{1} = \bar{R} + \delta R_{1} - \frac{\bar{R}}{2}\frac{\partial \bar{R}}{\partial z},
         \label{r1}     \end{equation}
          \begin{equation}
    R_{2} = \bar{R} + \delta R_{2} + \frac{\bar{R}}{2}\frac{\partial \bar{R}}{\partial z}.
         \label{r2}     \end{equation}
         Then
    \begin{equation}
  \textbf{d}_{g}  \equiv \left \langle \textbf{i}_{z} d_{0}\left (\frac{\bar{R}^7_{0}}{R_{1}^7}-\frac{\bar{R}_{0}^{7}}{R_{2}^7}
    \right ) \right \rangle = \textbf{i}_{z} d_{0}\left (\langle f_{1}^{-7} -  f_{2}^{-7}\rangle
   + \langle f_{1}^{-8}+ f_{2}^{-8}\rangle  \frac{7}{2}\frac{\partial \bar{R}}{\partial z}
    \right ),
         \label{d0-1}     \end{equation}
     \begin{equation}
    f_{j}= 1+ \delta R_{j}/\bar{R}.
         \label{fj}     \end{equation}
         With regard for $\langle f_{1}^{-J}\rangle = \langle
         f_{2}^{-J}\rangle$   and (\ref{hd8}), we obtain
         \begin{equation}
  \textbf{d}_{g}   \approx  -\frac{7d_{0}\bar{R}\textbf{g}}{3u_{1}^{2}}\langle
  f_{1}^{-8}\rangle ,
         \label{d0-2}     \end{equation}
         where
   \begin{equation}
   \langle f_{1}^{-8}\rangle \approx  1+ \frac{9!}{7!2!}\langle \left (\frac{\delta R}{\bar{R}} \right )^{2}\rangle
    +  \frac{11!}{7!4!}\langle \left (\frac{\delta R}{\bar{R}} \right )^{4}\rangle +
     \frac{13!}{7!6!}\langle \left (\frac{\delta R}{\bar{R}} \right )^{6}\rangle +\ldots \approx
     2.3.
         \label{fj2}     \end{equation}
   Above, we used $\langle (\delta R)^{2J+1}\rangle = 0$ and assumed $\langle (\delta R)^{2J}\rangle = \langle (\delta
     R)^{2}\rangle^{J}$. The exact formula for the DM, which arose in
     an atom with the coordinate $z=0$ due to the presence of $\nabla_{z}n$ or
     $\nabla_{z}T$ in the medium, is as follows:
  \begin{equation}
  \textbf{d}   =   -\textbf{i}_{z}d_{0}\int\limits_{z>0}n(\textbf{r})g_{c}(\textbf{r})
   \cos{\theta}\frac{\bar{R}_{0}^{7}}{r^{7}}d\textbf{r} +
    \textbf{i}_{z}d_{0}\int\limits_{z<0}n(\textbf{r})g_{c}(\textbf{r})
   \cos{\theta}\frac{\bar{R}_{0}^{7}}{r^{7}}d\textbf{r}.
         \label{d-int}     \end{equation}
      It is necessary to bear in mind that $n$ and $T$ can vary in the
     corresponding half-space ($z>0$  or $z<0$). Therefore, $g_{c}$
     depends on $r$ and $z$.
     Since $g_{c}$ has maximum at $r \approx \bar{R}$, and the main contribution ($\sim 90\%$) to integrals (\ref{d-int})
       is given by $g_{c}$ at $r \lsim 1.5\bar{R}$, we estimate $\textbf{d}$, by assuming that values of $n$ and $T$ in the whole upper
     half-space
     are equal to those at $z=\bar{R}$ ($z=-\bar{R}$ for the lower half-space).
      In this case, $g_{c}(r,z)\equiv g_{c}(r)$, where $g_{c}(r)$ corresponds to the indicated $n$ and $T$ which are different
      in the upper and lower half-spaces.  Then
       \begin{equation}
  \textbf{d}   =   -\frac{\textbf{i}_{z}d_{0}}{4n_{0}} (nS_{7})|^{z=\bar{R}}_{z=-\bar{R}},
            \label{d-int2}     \end{equation}
    \begin{equation}
  S_{j}  =  \int\limits_{\Omega =4\pi} n_{0}g_{c}(r)\frac{\bar{R}_{0}^{j}}{r^{j}}d\textbf{r}, \ \ \ n_{0}=
  \bar{R}_{0}^{-3}.
            \label{s7}     \end{equation}
   Since $S_{7}=S_{7}(T,n)$, we obtain
     \begin{equation}
  \textbf{d}   =  \textbf{d}_{T} + \textbf{d}_{\rho},
            \label{d-int3}     \end{equation}
    \begin{equation}
  \textbf{d}_{\rho} =  -\frac{S_{7}d_{0}\bar{R}\nabla_{z}n}{2n}\left
   (1+ \frac{n}{S_{7}}\frac{\partial S_{7}}{\partial n} \right ),
                \label{dro}     \end{equation}
             \begin{equation}
  \textbf{d}_{T}   =
    -\frac{d_{0}\bar{R}}{2}\frac{\partial S_{7}}{\partial T}\nabla_{z}T.
            \label{dT}     \end{equation}
     The dependence of the structural factor $S(k)$ on the temperature
     is determined by the formula \cite{feenb}
   \begin{equation}
  S(k,T)  =  S(k,T_{0})\coth{(E/2T)}/\coth{(E/2T_{0})}.
            \label{sk}     \end{equation}
    In order to calculate $S_{7}(T)$, it is necessary to know the function $g_{c}(r,T)$. It can be found from $S(k,T)$
    with the help of the well-known equation
   \begin{equation}
  S(k)  =  1+n\int (g_{c}(r)-1)e^{-i\textbf{k}\textbf{r}}d\textbf{r}.
            \label{gc}     \end{equation}
   The numerical calculation of $S_{7}$  by (\ref{s7}) was carried out at
   temperatures from $0$ to $T_{\lambda}$,
   and the experimental function $S(k,T_{0}=1\,K)$ from \cite{ss} was chosen as a ``bare'' one, $S(k,T_{0})$.
    It turned out that $S_{7}$ depends very weakly on the temperature, so that
    $\partial S_{7}/\partial T = 6 \cdot
    10^{-4}K^{-1}$ with a high accuracy for $S_{7}\approx 14.879$ in the interval $T=1\,K \div T_{\lambda}$. This determined
    the value of $\textbf{d}_{T}$ (\ref{dT}).

      Consider $\textbf{d}_{\rho}$ (\ref{dro}), by setting $\nabla T=0$.
      As a consequence of the relation
      $g_{c}(n_{2},r)=g_{c}(n_{1},(n_{2}/n_{1})^{1/3}r)$ we obtain
       \begin{equation}
  S_{j}(n_{2})  =  \int\limits_{4\pi} n_{0}g_{c}\left (n_{1},\left (\frac{n_{2}}{n_{1}}\right )^{1/3}r\right )
  \frac{\bar{R}_{0}^{j}}{r^{j}}d\textbf{r} =
  \left (\frac{n_{2}}{n_{1}}\right )^{j/3-1}S_{j}(n_{1}),
            \label{s7n1}     \end{equation}
     \begin{equation}
  \frac{\partial S_{7}}{\partial n}  =  \frac{4S_{7}}{3n}.
            \label{s7n2}     \end{equation}
   Taking into account (\ref{dro}), we find
     \begin{equation}
  \textbf{d}_{\rho} \approx  -\frac{7S_{7}d_{0}\bar{R}\nabla_{z}n}{6n}.
                \label{s7n3}     \end{equation}
     Therefore, we obtain with regard for (\ref{hd6}) that the DM of a helium atom
    induced by the gravity force is
     \begin{equation}
  \textbf{d}_{g}   \approx   -\frac{7S_{7}d_{0}\bar{R}\textbf{g}}{6u_{1}^{2}},
            \label{d0-3}     \end{equation}
             which agrees with (\ref{d0-2}) and (\ref{fj2}).
      Respectively, the polarization of He-II due to the gravitational field is
      \begin{equation}
  \textbf{P}_{g} =  n\textbf{d}_{g}  \simeq  -\frac{7S_{7}d_{0}\nabla\rho}{6\bar{R}^{2}\rho} = \gamma_{\rho}\textbf{g}, \quad
  \gamma_{\rho} \approx  -\frac{7S_{7}d_{0}}{6\bar{R}^{2}u_{1}^{2}}.
         \label{Pg}     \end{equation}

          The relation $\textbf{d}_{\rho} = K\nabla\rho/\rho$ can be obtained
      also \cite{vol-ch} on the basis of formulas of work \cite{vol84}. In this case,
      the coefficient $K$  corresponds approximately to (\ref{s7n3}).
        The formula close to (\ref{s7n3}) was obtained in \cite{nac08}.

     A different result was found in work \cite{mel} on the basis of classical
    ``inertial'' mechanism:
           \begin{equation}
 \textbf{P}_{g}  \equiv n\textbf{d}_{g,i} = \gamma_{i}\textbf{g}, \ \ \ \gamma_{i} \approx
 \frac{\varepsilon- 1}{4\pi}\frac{m}{2Z|e|},
  \label{pol2} \end{equation}
    where $Z|e|$ is the charge of the atom nucleus ($-2e$ for $He^4$), $\varepsilon$ is the
    dielectric permittivity ($1.057$ for He-II), $m$ is the mass of an atom,
     the index $i$ means ``inertial''.  According to \cite{mel},
     the gravitational field is a source of the polarization.
       The relation (\ref{pol2}) is similar to the founded above formula
     (\ref{Pg}).     But, at $u_{1}=240\mbox{m}/\mbox{s}$, the value of $\gamma_{\rho}$ is 136 times as large as the value of
    $\gamma_{i}$ (\ref{pol2}) and has the different sign.
         Moreover, we should like to emphasize a more important point: according to the assumption made in
    \cite{mel}, polarization (\ref{pol2})
     is induced directly by the gravitational field, whereas the polarization (\ref{Pg})
    is induced by the concentration gradient.
        These are the different sources of the effect.  We
       will  compare  both of the approaches in more detail below in Sec. 4.

                  It is worth noting that analogous equations connecting $\textbf{P}$
        with  $\nabla \rho$ and $\nabla T$, follow from the equations for a nonsuperfluid liquid.
         Hence, the specificity of superfluidity is not manifested there.

          Formula (\ref{Pg})  determines the polarization of He-II by the gravitational field (the gravielectric effect).
      It turns out that this polarization is measurable. Let He-II
      be positioned in a cylindrical dielectric vessel characterized by the vertical axis,  height $L$, and
      radius
      $R$. Let the vessel be completely filled with helium. Let the $Z$ be directed upward.
The potential difference ${\sss \triangle}\varphi$
      between point A at the top of the vessel at the center of the surface of helium and
     point B on the bottom at the center of the base of the vessel can be found analogously to \cite{lt1} (by considering
      that the dipoles are positioned only in the volume of helium and by neglecting the polarization of the vessel):
        \begin{eqnarray}
    {\sss \triangle}\varphi &= & \varphi_A - \varphi_B = -2\varphi_B = 2\int\limits_{0}^{L} dz\int\limits_{0}^{R}
   \varrho d\varrho \int\limits_{0}^{2\pi} d\phi
   \frac{P_{z}(z) z}{\varepsilon_{He}(\varrho^2 +
   z^2)^{3/2}}
        = \nonumber \\   &=&
    \frac{4\pi P_{z}}{\varepsilon_{He}}(L+R-\sqrt{L^2 + R^2}), \
      P_{z} = -\gamma_{\rho}g.
       \label{u}  \end{eqnarray}
     In experiments, the sizes of a vessel were approximately as follows: $L=20\,\mbox{cm}$,
     $R=L/2$. With these values and (\ref{u}), we get
     $|{\sss \triangle}\varphi|\approx 57\,\mbox{nV}$. The real $|{\sss \triangle}\varphi|$ will be somewhat less due to the polarization of walls.
      It is a very small voltage, but we indicate that a less
     voltage, ${\sss \triangle}\varphi=10\,\mbox{nV}$, was measured  in works \cite{r1,r2}. However, it was alternating.

    In the analysis, we used implicitly the assumption
    that, on the microlevel, the atoms ``feel'' the
    density gradient. In the formulas obtained above,
    sometimes $\bar{R}$ was the average over the time, $\langle R(z)\rangle_{t}$
and sometimes --- over atoms, $(V/N)^{1/3}$.
     The atoms in helium are moving chaotically, and the question arises whether
these averages are equal to each other, in particular, at
    $\partial\rho/\partial z\neq 0$. That is, is the relation
   \begin{equation}
    \langle R(z)\rangle_{t} \equiv \bar{R}(z) = (V_{z}/N_{z})^{1/3}
         \label{sr}     \end{equation}
    true? Here, $V_{z}$ is the volume of a thin layer with the given $z$, and $N_{z}$ is the number of atoms in the layer.
    If (\ref{sr}) is satisfied, then the atoms ``feel'', on the microlevel, the density gradient,
    and formulas (\ref{r1}),  (\ref{d-int2}) and those following from them are valid.
         If (\ref{sr}) would not be satisfied, then $\langle
         R(z)\rangle_{t}$ would be different for different atoms even along a layer
         $z=const$, where the density is constant.
     However, the total $N$-particle wave function of helium-II
     is symmetric relative to a permutation of atoms both in the ground state and in the presence of
quasiparticles \cite{fey}. For $\partial\rho/\partial z\neq 0$,
     the symmetry is conserved in the plane $z=const$. Therefore, $\langle
         R(z)\rangle_{t}$ for different atoms of a layer
         $z=const$ must be equal to one another and, hence, to the instantaneous value of the
         average over atoms
         $(V_{z}/N_{z})^{1/3}$. That is, (\ref{sr})
         is valid.

          \section{POLARIZATION OF He-II BY  A SECOND-SOUND WAVE}
  A standing second-sound half-wave
         \begin{equation}
  T = T_0 - 0.5{\sss \triangle} T \cos(\omega_{2} t)\cos(z\pi/L)
    \label{t9}     \end{equation}
   was realized in experiments in \cite{r1}. Here, $L=\lambda_{2}/2$ is the length of a resonator, and
    $\omega_2$ is the second-sound frequency.
   A second-sound wave induces the variable gradient of $T$ in helium by (\ref{t9}), which leads to the appearance
   of the variable gradient of density. The latter can be determined
    from the coefficient of thermal
  expansion $\alpha_{T}$ by the formula \cite{pat,xal}
   \begin{equation}
  \nabla \rho =  -\rho\alpha_{T}\nabla T
    (1-u_{2}^{2}/u_{1}^{2})^{-1}.
        \label{t9b}     \end{equation}
     Relations (\ref{d-int3})--(\ref{dT}) and (\ref{t9b}) together with $u_{2}\ll u_{1}$  allow one to determine
both the mean DM arising in every atom of He-II due to
    the gradients of $\rho$ and $T$,
      \begin{equation}
  \textbf{d} = \textbf{d}_{\rho} + \textbf{d}_{T} \approx 0.5d_{0}\bar{R} \nabla T \left
  (7S_{7}\alpha_{T}/3-\partial S_{7}/\partial T \right ),
        \label{t11}     \end{equation}
and the polarization of He-II,
         \begin{equation}
  \textbf{P} = \textbf{P}_{\rho} + \textbf{P}_{T} \equiv n\textbf{d}_{\rho} + n\textbf{d}_{T}\approx
  0.5d_{0}\bar{R}^{-2} \nabla T\left (7S_{7}\alpha_{T}/3-\partial S_{7}/\partial T \right ).
        \label{t13}     \end{equation}
      At  $T=1.3 \div 2\,K$, we have $\partial S_{7}/\partial T =0.6\cdot
        10^{-3}K^{-1}$, which is much less than $-7S_{7}\alpha_{T}/3 = (21.9 \div
        415)10^{-3}K^{-1}$ (by the data on $\alpha_{T}$ \cite{pat}). Therefore, the quantity  $\partial S_{7}/\partial
        T$ in (\ref{t13}) can be omitted. Thus, the polarization in a second-sound wave due to $\nabla T$
        is much less than the polarization due to $\nabla \rho$.

                For the experiment performed in \cite{r1}, we obtain that the potential difference $U$ between the
        electrode and the ground is
        analogously  \cite{lt1}:
         \begin{equation}
      U(T) =\eta^{*}(T)\gamma_{b}(R/L) \cos(\omega_{2}t),
        \label{t14-0}     \end{equation}
        \begin{equation}
      {\sss \triangle} \varphi(T) \equiv 2\eta^{*}(T)\gamma_{b}(R/L) =
      |-7S_{7}\alpha_{T}/3+\partial S_{7}/\partial T |\frac{\pi \gamma_{b}(R/L)d_{0}}{2\varepsilon_{He}\bar{R}^{2}}{\sss \triangle} T ,
        \label{t14}     \end{equation}
      where ${\sss \triangle} \varphi$ is the amplitude of oscillations of $U$.
    At $T=1.3\,K$, we have $\frac{{\sss \triangle} \varphi}{{\sss \triangle} T} \approx
    \frac{\gamma_{b}(R/L)}{60}\frac{k_{B}}{2e}$, whereas the experiment
    gives $\frac{{\sss \triangle} \varphi}{{\sss \triangle} T} \approx
    \frac{k_{B}}{2e}$. Here, the factor $\gamma_{b}(R/L)\equiv \gamma_{bound}$
    is not related to the polarization (i.e., not to the relation $\textbf{P} =
       -\gamma\textbf{w}$), but it is determined by boundary conditions (see \cite{lt1}).
    For a short resonator \cite{r1}, $\gamma_{b} \simeq 1.38$; therefore $\frac{{\sss \triangle} \varphi}{{\sss \triangle} T} \approx
    \frac{1}{45}\frac{k_{B}}{2e}$, which takes $1/45$ of the experimental value. In the case of a long resonator, $\gamma_{b} \simeq 0.05$, and
    $\frac{{\sss \triangle} \varphi}{{\sss \triangle} T} \approx
    \frac{1}{1200}\frac{k_{B}}{2e}$.  At $T=1.8\,K$, we have
    $\frac{{\sss \triangle} \varphi}{{\sss \triangle} T} \approx
    \frac{1}{4.5}\frac{k_{B}}{2e}$ and $\frac{{\sss \triangle} \varphi}{{\sss \triangle} T} \approx
    \frac{1}{120}\frac{k_{B}}{2e}$ for the short and long resonators, respectively.

    As $T$ increases, the theoretical quantity ${\sss \triangle}
    \varphi/{\sss \triangle} T$ grows like $ \alpha_{T}$, whereas the experimental one ${\sss \triangle}
    \varphi/{\sss \triangle} T$  does not depend on $T$ and  the ratio $R/L$ of the sizes
    of a resonator. Thus, the above-presented bulk mechanism predicts a weak electric signal
 which is by
     1-3 orders lower than the observed signal
    and strongly depends on the sizes of a resonator and the temperature. This
    signal can be measured under a higher accuracy of measurements.

         We note two points related to the electric signal arising due to the volume polarization of He-II.
     First, such a signal must obligatorily depend on
     the sizes of a resonator. Moreover, this dependence would be quite strong due to the
     factor $\gamma_{b}$.
           Indeed, the volume polarization of helium is a result of the polarization of separate atoms, since helium contains nothing except for atoms.
           The atoms possess a tidal DM. Therefore, since the polarization is determined by the sum of atomic DMs, the factor
           $\gamma_{b}$ is sure to appear, and the signal ${\sss \triangle} \varphi$ would turn out to be strongly dependent on the length of a resonator, which was not observed in experiments.
      If the temperature
      of the whole resonator would be noticeably changes synchronously with second sound,
       this would induce oscillations of $T$ in helium outside of the resonator and the corresponding polarization.
        In this case, ${\sss \triangle} \varphi$
      would be formed in the volume of $\sim L^3$, and, hence, $\gamma_{b}$ and ${\sss \triangle}
      \varphi$ would be close to $\gamma_{b}$ and ${\sss \triangle}
      \varphi$ for
      a short resonator. That is, ${\sss \triangle}
      \varphi$ would be weakly dependent on the sizes of a resonator.
       But the experiment \cite{r3} indicates that the oscillations of $T$ in a second-sound wave
        do not enter practically into helium outside of a resonator due to thick walls and a low
       heat conduction of a resonator.

          The second point consists in that
          the term $\sim \partial S_{7}/\partial T$ in formulas (\ref{t13}) and (\ref{t14})
describes the polarization due to
          $\nabla T$  (at $\nabla \rho =0$). Since $S_{7}$ is expressed through the structural factor,
          this term involves all possible
          microscopic mechanisms leading to the volume polarization due to
          $\nabla T$. But the contribution of this term to the polarization is
          small. Therefore, such mechanisms  cannot explain
          the observed signal ${\sss \triangle} \varphi$.

            Thus, the electric signal ${\sss \triangle}
      \varphi$ induced by a second-sound wave must consist of two parts: the major one, ${\sss \triangle}\varphi_{S}$, and the
      minor one, ${\sss \triangle}\varphi_{V}$.
      The signal ${\sss \triangle}\varphi_{S}$ observed
      in the experiment \cite{r1} was apparently caused by some surface effect in helium or by
the gradient thermoemf arising in the electrode due to a gradient of temperatures. The quantity ${\sss \triangle}\varphi_{S}$
      is independent of the sizes of a resonator and $T$ and is a response only to
variations of the temperature. But the quantity ${\sss
      \triangle}\varphi_{V}$ is the above-discussed signal which has not been yet discovered experimentally.
It appears due to
      the volume polarization of He-II and depends strongly on the sizes of a resonator. The shorter a resonator, the stronger this effect.
        Therefore, it is easier to discover the effect with
      short resonators (where $L\lsim R$), and the resonator must be fabricated of a dielectric
      (in order to avoid induced noises and to increase $\gamma_{b}$).

         \section{CONNECTION BETWEEN ACCELERATION AND POLARIZATION}

      One of the goals of the present work is the study of the connection between the polarization and the acceleration for the atom or medium.
      This problem was considered in the work \cite{mel}, in which
      it was   found  that the polarization $\textbf{P}$
    of a dielectric is proportional to its acceleration $\textbf{w}$:
       \begin{equation}
 \textbf{P}_{i} = -\gamma_{i}\textbf{w}
  \label{pol1} \end{equation}
  with $\gamma_{i}$ from (\ref{pol2}). According to \cite{mel},
     the acceleration $\textbf{w}$ is a source of the
     polarization.
           However, the reasoning underlying formula
     (\ref{pol1}) is not quite correct, in our opinion. It was assumed in the derivation of
     (\ref{pol1}) that the electron shell of a single atom of the quiescent dielectric is stretched by the gravitational field.
      But the gravity leads to the identical acceleration for electrons and the nucleus. Therefore,  it cannot stretch any atom.
       Hence, the reasoning in \cite{mel} and formula (\ref{pol1}) lose
       the significance.
      In addition,  in \cite{mel}, the interaction of a single atom with the other atoms
      was considered only in terms of $\varepsilon$, i.e. on the macroscopic level; but such a situation should be analyzed
      microscopically, by explicitly taking the interaction of the atom with its neighbors into account.

      Let us assume that the gravity does stretch an atom. We will show that, in this case, the explicit
      consideration of its interaction with the neighboring atoms is of importance.
       According to \cite{mel},   a quiescent dielectric in a gravitational
       field acquires  the polarization (\ref{pol2}), which yields (\ref{pol1}).  The brief substantiation of
    formula (\ref{pol2}) can be found in \cite{mel} and, in more details, in \cite{naz}. The reasoning of works \cite{mel,naz}
    is as follows. The nucleus of an atom of the dielectric undergoes the action of the gravity force
    $m_{c}\textbf{g}$ and somewhat
    ``sags'' by the distance $\delta_{g}$. As a result, the atom acquires a DM of $Z|e|{\bf\delta}_{g}$,
    and the dielectric becomes polarized. For small deformations ${\bf\delta}_{g}=\eta
    m_{c}\textbf{g}$, $\eta$ is determined from the following relations for an atom in an
    electric field: $d=\kappa E/n$,
    $d=Ze\delta_{E}=Ze\eta(2ZeE)$, where $\kappa = (\varepsilon - 1)/4\pi$. From this relations, we get $\eta =
    \frac{\kappa}{2n(Ze)^2}$ and formula (\ref{pol2}). In such an approach,
    the gravitational field acting on the atom nucleus,
    is compensated by the force $\textbf{F}_{d}$ related to the elastic deformation of the electron
    shell of the atom:
      \begin{equation}
     \textbf{F}_{d}+m_{c}\textbf{g}=0, \quad \textbf{F}_{d}=
    -{\bf\delta}_{g}/\eta.
      \label{rav} \end{equation}

        Let us consider the problem in more details.
         We can easily evaluate the ratio of the forces affecting an atom in a dielectric, by considering, firstly, a dielectric
         without the gravity and then by
        ``switching-on'' the gravity force. Without the gravity, only the interatomic forces act on
        the atoms of a dielectric. From the side of each of the neighboring atoms, a specific atom
        is affected by: 1) the van der Waals force
        $\textbf{F}_{v}$ which is well-known at $R\gsim a$ for helium-II \cite{aziz}:
         \begin{equation}
    \textbf{F}_{v} = - \nabla U, \ \ U(R) = 4\varepsilon\left (\left (\frac{a}{R} \right )^{12} -
   \left (\frac{a}{R} \right )^{6} \right ), \quad
   \varepsilon=11K, \ a=2.64\AA,
    \label{lj} \end{equation}
      where $R$ is the interatomic distance, $a$ is the ``size'' of an atom;
     2) the force of ``quantum pressure'' (arising due to the motion of atoms caused by zero-point
     oscillations which increase the mean energy
     of an atom  at  $T=0$ from $-\varepsilon = -11\,K$ to
      $E_{0} = -7.16\,K$)  equal approximately to $F_{qp}\simeq {\sss \triangle} p/({\sss \triangle} t)
      \simeq 2\langle p \rangle/({\sss \triangle} t) \simeq \langle p
      \rangle^{2}/(m_{4}(R-a)) \simeq \hbar^{2}/(m_{4}(R-a)^3)$;
      3) the pressure due to the thermal motion of atoms.

         Let us switch-on the gravity force. It gives
         the same acceleration to electrons and the nucleus. So,
             each atom of a dielectric starts to move downward, and the dielectric will
         be contracted.  He-II has a high
         heat conductivity; therefore, the temperature and the thermal pressure at all points
         of a vessel will be fast equalized. In this case, the gravity force
         must be compensated owing to that the interatomic force
         $\textbf{F}_{v}+\textbf{F}_{qp}$ acting on every atom from the side of lower neighboring atoms
         exceeds the analogous force from the side of upper neighboring atoms
         by $m_{4}g$. To realize this, the gradient density should be present in a dielectric.
           How can a DM of helium atoms be determined in this case? In
           \cite{mel,naz}, the DM was determined on the basis of the assumption that
the atom is elongated in the gravitational field by $\delta_{g}$
so that the gravity force acting on the nucleus is compensated by
the elasticity of an atom according to (\ref{rav}).
                      We now study whether the atom will be elongated in such a way.
           To this end, we compare the elastic force
          with interatomic forces.
           Let the atom move downward by a distance of $\delta_{g}$. This results in the appearance of the difference of the distances
to the neighboring upper and lower atoms
           $\delta\bar{R}=2\delta_{g}$, as well as the difference of the van der Waals forces, $\delta
           F_{v}=F_{v}(\bar{R})-F_{v}(\bar{R}-2\delta_{g})=
           F_{v}^{\prime}(\bar{R})2\delta_{g}= -
           U^{\prime\prime}(\bar{R})2\delta_{g}$.
           At the tension by $\delta_{g}$, the elastic force
           is $F_{d}=m_{c}g\approx m_{4}g$ according to (\ref{rav}).
           In view of (\ref{lj}), we obtain
     \begin{eqnarray}
    \frac{\delta F_{v}}{F_{d}}&=& - \frac{U^{\prime\prime}(\bar{R})2\delta_{g}}{m_{4}g}=-2\eta U^{\prime\prime}(\bar{R})
     =  \nonumber \\
     &=& -\frac{8\varepsilon\eta}{\bar{R}^2}\left (\left (\frac{a}{\bar{R}} \right )^{12}12\cdot 13 -
   \left (\frac{a}{\bar{R}} \right )^{6} 6\cdot 7 \right )\approx 2.9\cdot 10^{-6},
    \label{lj2} \end{eqnarray}
      where $\bar{R}=\bar{R}_{0} = 3.578\,\mbox{\AA}$ is the mean distance
      between atoms of He-II at $\rho=0.1452 \mbox{g}/\mbox{cm}^3$.
       That is, the difference of the van der Waals forces arisen due to the lowering
       of an atom at a distance of $\delta_{g}$ is \textsl{by 6 orders less} than the
       forces owing to the atomic elasticity appearing at the tension of an atom
       by the same $\delta_{g}$. This implies that the atom is a
       very elastic object and will fall down without tension. In the above estimates, we
       neglect the quantum pressure, whose contribution is of the same order as that of the van der Waals force,
but its correct modeling is a difficult task.

                 Thus, the atom in a gravitational field must not be stretched by
         $\delta_{g}$, because the gravity force affecting the nucleus is compensated by the
difference of interatomic forces for the adjacent layers of atoms,
rather than by the tension of an atom.
          A slight tension is possible, but it will lead to a polarization significantly less than
         (\ref{pol2}), because we have $\delta_{g}/\eta \ll m_{c}g$ instead of $\delta_{g}/\eta=m_{c}g$.
          Therefore, $\gamma_{i}$ in formulas (\ref{pol1}) and (\ref{pol2})
          describing the inertial polarization related to the elastic tension of an atom
                    must be significantly less than $\gamma_{i}$ given in (\ref{pol1}).
            While studying the accelerated motion of a dielectric, we need also to consider the interaction of atoms,
             because the inertia forces do not exist in the nature
         and are used only as a convenient mean to describe the accelerated motion of bodies caused by the real forces
         (for the atoms of He-II, they are interatomic forces).

          The estimates indicate that the inertial mechanism of the polarization
          must take the interaction of atoms into account and, hence,
          must be developed in the frame of quantum theory.
          Let us consider a single atom. To move with acceleration, it should
          undergo the action of an external force, e.g.,
          the gravity force or the electromagnetic wave.
          In both cases, the atom will not be polarized by acceleration,
          because the inertial forces (and the gravity force) accelerate electrons and the atomic nucleus identically.
           But the source of a force can be another atom. This is a partial case with a nonzero density gradient where
           two immovable He atoms polarize each other according to (\ref{da}).
       Let the distance between the atoms be somewhat larger than the equilibrium one
       (corresponding to the  minimum of the potential (\ref{lj})). In this case, the equilibrium does not hold, and the atoms
       begin to approach each other.
        As a result, the tidal DM of each atom is changed. By expanding the DM in a series
in the small displacement $\delta R$, velocity, acceleration $w$,
etc., we can evaluate the contribution of these parameters to the
         DM of an atom with the help of nonstationary
         perturbation theory. In this case,  the acceleration is the
         source of the corresponding correction to the DM, and it is natural to name this correction
         the inertial polarization of an atom.
         The preliminary analysis showed that it $\sim
         w^2$. That is, this correction is small, and the zero approximation (\ref{da}) (for immovable atoms)
          gives the main contribution. In the derivation
         of formulas (\ref{dro}) and (\ref{dT}), the atoms were considered
         immovable.  We did not calculate the inertial correction to a macroscopic polarization
         and \textsl{consider the question about the value of $\gamma_{i}$ in (\ref{pol1})
         to be open}. Most probably, $|\gamma_{i}| \ll
         |\gamma_{\rho}|$, but we do not exclude completely  the possibility that $|\gamma_{i}| \gsim
         |\gamma_{\rho}|$.

         In the zero approximation (i.e. for immovable atoms),
         the connection between the polarization $\textbf{P}$ and
     the  acceleration $\textbf{w}$ of the medium can be determined from
     formulas  (\ref{Pg}) and (\ref{dT}). According these formulas, if
     $\nabla \rho$ and/or $\nabla T$ are nonzero in the medium, the medium
     is polarized.  It is obvious from (\ref{hd2}) and (\ref{hd5}) that if the gravity force
   is absent, but there is the constant gradient of density
   in He-II (\ref{hd6}), then helium as a whole must move with
   the acceleration $\textbf{w}=-\textbf{g}$. In the reference system where helium is at rest,
   relation (\ref{Pg}) holds. It allows us to get the connection between the polarization of helium and its
   acceleration:
     \begin{equation}
  \textbf{P}_{\rho} = -\gamma_{\rho} \textbf{w}_{\rho}, \ \
  \gamma_{\rho} \simeq -\frac{7S_{7}d_{0}}{6\bar{R}^{2}u_{1}^{2}}.
         \label{Pr}     \end{equation}
          We now determine the connection of the polarization
        $\textbf{P}_{T}$       appeared due to the gradient of $T$ with the
        corresponding acceleration $\textbf{w}_{T}$.
        It is seen from the equations
        \begin{equation}
 \rho \textbf{w} \equiv \rho D\textbf{v}/Dt=-\nabla p,
  \label{t15} \end{equation}
              \begin{equation}
  \nabla p =  \frac{\partial p}{\partial \rho}|_T \nabla \rho +
   \frac{\partial p}{\partial T}|_{\rho} \nabla T \approx
     u_{1}^{2}\nabla \rho + \frac{\partial (SV)}{\partial V}|_{T} \nabla
     T,
  \label{t16} \end{equation}
      \begin{eqnarray}
  \frac{\partial (SV)}{\partial V}|_T &=& -\frac{\rho}{V}\frac{\partial (SV)}{\partial
  \rho}|_T = -\frac{\rho}{V}\frac{\partial (SV)}{\partial p}|_T
  \frac{\partial p}{\partial\rho}|_{T} \approx \nonumber \\
  &\approx & \frac{\rho u_{1}^{2}}{V}\frac{\partial^{2} (FV)}{\partial p \partial
  T} = \frac{\rho u_{1}^{2}}{V}\frac{\partial V}{\partial T}|_{p}
  = \rho u_{1}^{2}\alpha_{T}
  \label{t16-2} \end{eqnarray}
   that the acceleration of a fluid element is given by two terms
     which are related, respectively, to $\nabla \rho$ and $\nabla T$:
     \begin{equation}
   \textbf{w} \approx
     -u_{1}^{2}\nabla \rho/\rho - u_{1}^{2}\alpha_{T}\nabla T \equiv \textbf{w}_{\rho}+\textbf{w}_{T}.
  \label{t17} \end{equation}
   Relations (\ref{t17}) and (\ref{dro}) yield  (\ref{Pr}), and relations (\ref{t17}) and (\ref{dT})
   allow us to get
       \begin{equation}
    \textbf{P}_{T}=-\gamma_{T}\textbf{w}_{T}, \quad \gamma_{T} \approx
           \frac{\partial S_{7}}{\partial T}
           \frac{d_{0}}{2u_{1}^{2}\alpha_{T}\bar{R}^{2}}.
              \label{t20}     \end{equation}
       In this case, $\gamma_{\rho} \simeq -136\gamma_{i}$ and   $\gamma_{T}(T=1.3\div 2\,K) \simeq (0.017 \div
         0.001)\gamma_{\rho}$.

         The acceleration  in a second-sound wave, according to
         relations (\ref{t9b}) and (\ref{t17}), is equal to
     $\textbf{w}_{2} \approx  - u_{2}^{2}\nabla \rho /\rho$  \cite{pat,xal}.
     That is, the accelerations induced by the gradients of
         $\rho$ and $T$ are almost the same by modulus and of the opposite signs.
            In this case, the polarizations $\textbf{P}_{\rho}$  and $\textbf{P}_{T}$ do not compensate each other,
         since $P_{T}\ll P_{\rho}$.   For a second-sound wave, the total polarization and acceleration are
         connected by the relation
     \begin{equation}
    \textbf{P}=\textbf{P}_{\rho} + \textbf{P}_{T} = -\gamma_{2}\textbf{w}_{2}, \quad
        \gamma_{2} \approx   (u_{1}/u_{2})^{2}\gamma_{\rho},
              \label{pw2}     \end{equation}
          so that  $\gamma_{2} \sim  10^{2}\gamma_{\rho} \sim -10^{4}\gamma_{i}$.

        It is seen from formulas
         (\ref{Pr}),  (\ref{t20}), and (\ref{pw2}) that there is no
         universal connection between $\textbf{P}$ and $\textbf{w}$
         in the medium,  and the
      value of $\gamma$ depends on the mechanism of the polarization.
     In addition, despite the fact that the polarization is proportional to the
      acceleration, the gradient $\nabla \rho$ or $\nabla T$ is the \textit{source} of a polarization, rather than
       the acceleration or the gravity field (\ref{Pg}). The acceleration
       is another consequence of $\nabla \rho$ (or  $\nabla T$).

       The tidal DMs and higher multipoles of atoms arising in all these cases
       induce the arbitrarily small forces (as compared with the
       van der Waals force) in the medium.  Therefore, the tidal polarization of atoms
       should not be included in the equations of two-fluid hydrodynamics;
        it does not affect $\nabla \rho$ or  $\nabla
       T$ and cannot compensate the gravity field.
      The tidal DM is similar to a vane which shows the direction of a wind (the direction to the nearest atom) but does not affect it.

    Thus, though the polarization of individual atoms is induced by the tidal mechanism
    and not by acceleration, however,
    at nonzero $\nabla\rho$ and/or $\nabla T$ the spatial location of the atoms
is such that instantaneous distances between them do not
correspond to the equilibrium ones, so that interaction between
the atoms leads to their acceleration $\textbf{w}$, and an average
polarization of the atoms turns out to be proportional to the
average value of $\textbf{w}$, as was first suggested in
\cite{mel}.

   Apparently, for any microscopic origin
      of $\textbf{w}$, there is the linear connection
      $\textbf{P}=-\gamma\textbf{w}$ if $\textbf{w}$ is small. Indeed, both the polarization and
      the acceleration are caused by the directed inhomogeneity of the
      position or motion of atoms, the values of $\textbf{P}$ and
      $\textbf{w}$ are proportional to this inhomogeneity and, hence, to each other.

       We should like to note that work
\cite{mel} became one of the stimuli for the present study.

                   \section{CONCLUSIONS}

             In the present work, the polarization of He-II due to the gradients of density and temperature is calculated,
             the relation of the polarization to the acceleration of the medium is studied, and the gravielectric effect is considered.
           It is shown that a second-sound wave should induce a volumetric
      electric signal ${\sss \triangle} \varphi$ mainly arising due to the gradient of $\rho$ in helium.
      This signal is essentially less than the registered one and can be observed within the future more exact measurements.
      As for the Rybalko's effect (i.e., the observed signal ${\sss \triangle} \varphi \simeq
    k_{B}{\sss \triangle} T/2e$), it is caused, apparently, by a surface mechanism.

       \section*{ACKNOWLEDGMENT}

The author is grateful to V.M.~Loktev, L.A.~Melnikovsky,
A.S.~Rybalko,
  and G.E.~Volovik for the helpful  discussions and remarks.
  The work is performed under the financial support
  of the Division of Physics and Astronomy of the NAS of Ukraine in the frame of a Target program
  of fundamental studies N 0107U000396.

\renewcommand\refname{REFERENCES}

       \end{document}